\documentclass{article}

\begin{document}

\title{Wave-like spatially homogeneous models\\
of St\"{a}ckel spacetimes (2.1) type\\
in the scalar-tensor theory of gravity}

\author{
Konstantin~Osetrin\thanks{osetrin@tspu.edu.ru},  \\[1ex]
Altair~Filippov\thanks{altair@tspu.edu.ru},  \\[1ex]
Evgeny~Osetrin\thanks{evgeny.osetrin@gmail.com}, \\[2ex]
Tomsk State Pedagogical University, \\[1ex]
Tomsk, 634061, Russia\\
}

\maketitle

\begin{abstract}
Six exact solutions are obtained in the general scalar-tensor theory of gravity related to spatially homogeneous wave-like models of the Universe. Wave-like space-time models allow the existence of privileged coordinate systems where the eikonal equation and the Hamilton-Jacobi equation of test particles can be integrated by the method of complete separation of variables with the separation of isotropic (wave) variables on which the space metric  depends (non-ignored variables).
An explicit form of the scalar field and two functions of the scalar field that are part of the general scalar-tensor theory of gravity are found.
The explicit form of the eikonal function and the action function for test particles in the considered models is given.
The obtained solutions are of type III according to the classification \mbox{Bianchi} and type N according to the classification of \mbox{Petrov}.
\end{abstract}

%\keywords{scalar-tensor theory of gravity; wave-like spacetimes; spatially homogeneous spacetimes; exact solutions.}

\section{Introduction}	

When searching for tools for a theoretical description of the phenomena of accelerated expansion of the Universe, dark matter and dark energy, modified gravity theories and, in particular, scalar-tensor gravity theories 
\cite{1}-\cite{3} are in the area of ​​interest of researchers. With the beginning of the era of gravitational wave astronomy, one of the possibilities for selecting viable modified theories of gravity may be the study of the space-time wave models allowed by them, including primordial gravitational waves in the Universe. Therefore, the study of wave-like spatially homogeneous models in Einstein's theory and in modified theories of gravity is of additional interest.

In connection with the foregoing, it is of interest to study spatially homogeneous space-time models that allow the existence of privileged coordinate systems (CS), where it is possible to separate variables in the Hamilton-Jacobi equation for test particles and in the eikonal equation with separation of isotropic (wave) variables. Such space-time models that allow the separation of isotropic (wave) variables will be called wave-like Shapovalov spaces (in honor of V.N.~Shapovalov, who first distinguished these classes of spacetimes \cite{4}-\cite{5}).

In this paper, we study wave-like spatially homogeneous space-time models that admit two non-ignored variables (which determine the metric of these models in a privileged CS) for the case of the general scalar-tensor theory of gravity. % These models allow two Killing commuting vectors.
Earlier, we obtained three types of such wave-like spatially homogeneous models with two commuting Killing vectors. 

Further in the article, we obtained exact solutions for these models in the general scalar-tensor theory of gravity in vacuum. The explicit form of the scalar field in the ''separated'' form is found, as well as the explicit form of the two functions of the scalar field -
$ \omega (\phi) $ and $ \Lambda (\phi) $, which are part of the general theory.

We consider a generalized scalar-tensor theory of gravity with a Lagrangian of the form
\begin{equation}
{\it L}=\frac{1}{4\pi}\sqrt{-g} \left[\phi R - \frac{\omega(\phi)}{\phi}\nabla_i\phi\nabla^i\phi-2\Lambda(\phi) \right]+
L_{matter}.
\label{Lagrang}
\end{equation}

Here the function $ \Lambda (\phi) $ plays the role of a generalized cosmological constant. Note that the general scalar-tensor theory transforms to the Brans-Dicke theory when $ \omega $ turns into a constant, and the functions $ \Lambda $ into $ \lambda \, \phi $, where $ \lambda $ is an ordinary cosmological constant.

Vacuum field equations for the theory with the Lagrangian (\ref{Lagrang}) take the form
\begin{equation}
\phi G_{ij}+\left( \nabla^k\nabla_k \phi+\frac 12\frac{\omega(\phi) }{\phi}\nabla^k\phi\nabla_k\phi+\Lambda(\phi)\right)g_{ij}-\nabla_i\nabla_j\phi-\frac{\omega(\phi) }{\phi}\nabla_i\phi\nabla_j\phi=0,
\label{EqField}
\end{equation}
the scalar field equation can be written as follows
\begin{equation}
\bigl(2\omega(\phi)+3 \bigr)\nabla^k\nabla_k \phi+\omega'(\phi)\nabla^k\phi\nabla_k\phi+4\Lambda(\phi)-2\phi \Lambda'(\phi)=0.
\label{EqScalar}
\end{equation}
Here $ G_{ij} = R_{ij} -1/2 \, R \, g_{ij} $ is the Einstein tensor, $ \nabla_i $ is the covariant derivative, the prime means the derivative with respect to the scalar field $ \phi $.

In this paper, we obtain exact solutions of the field equations of the generalized scalar-tensor theory of gravity 
(\ref{EqField})-(\ref{EqScalar}) for classes of wave-like spatially homogeneous space-time models that allow separation
isotropic (wave) variables in the eikonal equation
\begin{equation}
g^{ij}\,\partial_i\Psi\,\partial_j\Psi=0
\label{EqEikonal}
\end{equation}
and in the Hamilton-Jacobi equation for a test particle of mass $m$
\begin{equation}
g^{ij}\,\partial_i S\,\partial_j S=m^2,
\label{EqHJ}
\end{equation}
where $ \Psi $ is the eikonal function, $ S $ is the test particle action function, $ \partial_i $ is the partial derivative.

Spaces that allow the existence of privileged coordinate systems, where the equation of motion of test particles in the Hamilton-Jacobi form (\ref{EqHJ}) allows integration by the method of complete separation of variables called Stackel spaces (in honor of P.~St\"{a}ckel - see \cite{6, 7}) . A brief summary of the theory of Stackel spaces can be found in \cite{4, 5, 8}.

In this paper, we will consider the case when the metric in a privileged CS (where separation of variables is allowed) depends on two variables. The variables that the metric depends on in privileged coordinate systems are called non-ignored.
For the case of two non-ignored variables, there are three types of similar spaces -
{A4}, {B1} and {B2} (see \cite{9}-\cite{11}). Models of type A are a subclass of St\"{a}ckel spaces of type (2.1), and models of type B are conformal-St\"{a}ckel spaces of type (3.1).

The theory of St\"{a}ckel spaces is widely used in various metric theories of gravity, see f.e. \cite{Bagrov} - \cite{OO}.

\section{Spatially homogeneous wave-like model {A4} type  }

A model of a wave-like spatially homogeneous space-time of type {A4} in a privileged coordinate system (where the eikonal equation (\ref{EqEikonal}) allows separation of variables) has the following metric (see \cite{11}):
\begin{equation}
ds^2=\frac{1}{{x^0}^2}\,\left[
\,{dx^0}^2+2\,dx^1dx^3+
\frac{1}{{x^0}^2}\,\left(
dx^2-x^1\,dx^3
\right)^2
\,
\right],
\label{IntervalA4}
\end{equation}
where $ x^1 $ is an isotropic variable.

Independent Killing vector fields of model {A4} in a privileged CS can be selected in the form:
\[
X_0=\partial_3,
\qquad
X_1=\partial_2,
\qquad
X_2=-x^1\partial_1+x^3 \partial_3,
\]
\begin{equation}
X_3=x^0\partial_0+2{x^1}\partial_1+ 2 x^2\partial_2,
\end{equation}
Killing vectors $ X_1 $, $ X_2 $, $ X_3 $ define a subgroup of spatial homogeneity of the model.
Killing Vectors commutators of  {A4} model have the form:
\[
 [{X_0},{X_1}]=0,\qquad [{X_0},{X_2}]={X_0},\qquad [{X_0},{X_3}]=0,
 \]
\begin{equation} 
 [{X_1},{X_2}]=0,\qquad [{X_1},{X_3}]=2{X_1},\qquad [{X_2},{X_3}]=0.
 \end{equation}
 
Sign-definiteness of the metric on the orbits of the subgroup of spatial homogeneity imposes restrictions on the range of permissible values ​​of the coordinates used:
$ x^1x^3 <0, \ \ {x^0}^ 2> 2 \, | x^1x^3 | $.
 
The scalar curvature is constant and negative,
the Ricci tensor and the Weyl tensor do not vanish.
The model is of type ~ D according to Petrov’s classification and belongs to type~{III} according to Bianchi classification.

Notice, that
 solutions of the Einstein equations with pure radiation as a matter
for model {A4}
are absent.
An analysis of the field equations of the general theory of gravity (\ref{EqField})-(\ref{EqScalar}) for the model 
(\ref{IntervalA4}) also shows the absence of vacuum solutions in the scalar-tensor theory of gravity.

\section{Spatially homogeneous wave-like model  {B1} type}

The interval for the space-time model of type {B1} has the form (see \cite{10}):
\begin{equation}
ds^2=\frac{1}{{x^0}^2}\,\left[ 
{dx^0}^2+2\,dx^1dx^3+(x^1-\alpha)^{1-\beta}(x^1+\alpha)^{1+\beta}\,{dx^2}^2
\right],
%\quad
%\alpha\neq 0,\  \beta^2\neq 1,
\label{MetricB1}
\end{equation}
where $ x^1 $ is an isotropic (wave) variable, $\alpha$ and $\beta$ are constants ($\alpha\ne 0$, $\beta\ne\pm 1$).

Independent Killing vector fields of model {B1} in a privileged CS can be selected in the form:
\[
X_0=\partial_3,
\qquad
X_1=\partial_2,
\qquad
X_2=x^0\partial_0+ x^2\partial_2+ 2\,x^3 \partial_3,
\]
\begin{equation}
X_3=x^0 x^1\partial_0+({x^1}^2-\alpha^2)\partial_1+ {\alpha} \beta x^2\partial_2-\frac{{x^0}^2}{2}\partial_3.
\end{equation}
Killing vectors $ X_1 $, $ X_2 $, $ X_3 $ define a subgroup of spatial homogeneity of the model.

Killing Vectors commutators of  {B1} Model have the form:
\[
[{X_0},{X_1}]=0,
\qquad
[{X_0},{X_2}]=2{X_0},
\qquad
[{X_0},{X_3}]=0,
\]
\begin{equation}
[{X_1},{X_2}]={X_1},
\qquad
[{X_1},{X_3}]={\alpha}\beta{}{X_1},
\qquad
 [{X_2},{X_3}]=0.
\end{equation}
For $ \beta = 0 $, this space admits an additional Killing vector and degenerates into
space with one non-ignored variable in privileged CS.

The condition of spatial homogeneity of the model imposes restrictions on the region of admissible coordinate values.
It follows from the form of the metric determinant that $ | {x^1} |> | {\alpha} | $, and the spatial homogeneity of the model
Highlights the allowed range of variables in the used CS:
$ | \, {{x^0}^ 2 + 2 \, x^1x^3} \, | <4 \, | \, {\alpha} \, {x^3} \, | $.
 
 The scalar curvature of the model is constant and negative,
 the Ricci tensor does not vanish, the components of the Weyl tensor are proportional to the following expressions
  $ C_{ijkl} \sim {\alpha} \, ({\beta}^2-1). $

This spatially homogeneous space-time model is of type {III} according to the Bianchi classification and has type {N} according to the Petrov classification.

It is assumed that the scalar field in the preferred coordinate system has the ''separated'' form:
\[
\phi=\phi_0(x^0)\,\phi_1(x^1)\,\phi_2(x^2)\,\phi_3(x^3).
\]
The field equations of the general scalar-tensor theory of gravity in vacuum (\ref{EqField}) for the model {B1} 
with the metric (\ref{MetricB1}) can then be written as:
\[ 
 2 \Lambda\, \phi_0 \phi_2 ({x^1}^2-\alpha^2 ) = 
 \phi_0^2 \phi_1 \phi_3 \,\omega\,      {x^0}^2\phi_2{}'^2 ({x^1}-\alpha )^\beta 
 /({x^1}+\alpha )^\beta 
 -\mbox{}
\]
\begin{equation} 
\phi_2^2 ({x^1}^2-\alpha^2 )  
  \Bigl(\phi_1 \phi_3 (6 \phi_0^2 - 6 \phi_0 {x^0}         \phi_0{}' - \omega {x^0}^2 \phi_0{}'^2) +
2 \phi_0^2       (3 + 2 \omega) {x^0}^2       \phi_1{}' \phi_3{}'
\Bigr), 
\label{EqFieldB11}
\end{equation} 
\begin{equation}  
\phi_2{}''  =
%-\frac{\omega\phi_2{}'^2 }{\phi_2}+
\frac{ \phi_2 (\alpha + {x^1})^\beta    \left( \phi_1 (\alpha \beta - {x^1}) - (1 + \omega) 
      (\alpha^2 - {x^1}^2) \phi_1{}'\right) \phi_3{}'
}{  
\phi_1  \phi_3 (-\alpha + {x^1})^\beta }
-\frac{\omega\phi_2{}'^2 }{\phi_2} , 
\end{equation} 
\begin{equation}  \phi_0{}''  =  \frac{-2 \phi_0{}'}{{x^0}} -   \frac{\omega \phi_0{}'^2}{\phi_0} + \frac{\phi_0 \phi_1{}' \phi_3{}'}{\phi_1 \phi_3} + 
\frac{\phi_0 \omega     \phi_1{}' \phi_3{}'}{\phi_1 \phi_3}  , \end{equation} 
\[ 
\phi_1{}''  =  -\alpha^2 (-1 + \beta^2) \phi_1/(\alpha^2 - {x^1}^2)^2 -   \frac{\omega \phi_1{}'^2}{\phi_1} 
\] 
\begin{equation}  
\left(\phi_1 (-\alpha \beta + {x^1}) +    (1 + \omega) (\alpha^2 - {x^1}^2)     \phi_1{}'\right) \phi_2{}'=0  , 
\end{equation} 
\begin{equation}  (\phi_0 + (1 + \omega)     {x^0} \phi_0{}') \phi_1{}'=0  , \end{equation} 
\begin{equation}  (\phi_0 + (1 + \omega)     {x^0} \phi_0{}') \phi_2{}'=0  , \end{equation} 
\begin{equation}  (\phi_0 + (1 + \omega)     {x^0} \phi_0{}') \phi_3{}'=0  , \end{equation} 
\begin{equation}  \omega     \phi_3{}'^2 + \phi_3 \phi_3{}''=0  , \end{equation} 
\begin{equation}  (\omega+1)\phi_2{}'\phi_3{}'=0. 
\label{EqFieldB19}
\end{equation} 
The scalar equation (\ref{EqScalar}) for model {B1} takes the form:
\[
-2 \Lambda' \phi_0 \phi_1 \phi_2 \phi_3 +  4 \Lambda + \omega' \phi_1 \phi_3 {x^0}^2  
\Bigl[ 
2 \phi_0^2 \phi_2^2     \phi_1{}' \phi_3{}' +
\mbox{}
\]
\[
\phi_1 \phi_3 
\Bigl(
\phi_2^2 \phi_0{}'^2 +  
    \phi_0^2 (-\alpha + {x^1})^{-1 + \beta} (\alpha + {x^1})^{-1 - \beta}       \phi_2{}'^2
\Bigr)
\Bigr] + 
\mbox{}
\]
\[
\Bigl[
(3 + 2 \omega) {x^0} (\alpha + {x^1})^{-1 - \beta}   
\Bigl(
     \phi_2 (\alpha + {x^1})^\beta 
\bigl[
\,
2 \phi_0 {x^0} (\alpha^2 - {x^1}^2)        \phi_1{}' \phi_3{}' + 
\mbox{}
\]
\[  
\phi_1
\bigl(
\phi_0 {x^0} (\alpha \beta - {x^1}) \phi_3{}' +  
\phi_3 (\alpha^2 - {x^1}^2) (-2 \phi_0{}' +           {x^0} \phi_0{}'')
\bigr)
\,
\bigr]
\]
 \begin{equation}
 -         \phi_0 \phi_1 \phi_3 {x^0}      (-\alpha + {x^1})^\beta \phi_2{}''
\Bigl)
%)
\Bigr]
/(\alpha - {x^1})=0.
\label{EqScalarB1}
\end{equation} 
When integrating the field equations (\ref {EqFieldB11})-(\ref{EqScalarB1}) for the model {B1}, we obtain two types of exact solutions, which are given below.

\subsection{Exact solution \#1 of field equations for model {B1}}

\[
ds^2=\frac{1}{{x^0}^2}\,\left[ 
{dx^0}^2+2\,dx^1dx^3+(x^1-\alpha)^{1-\beta}(x^1+\alpha)^{1+\beta}\,{dx^2}^2
\right],
\]
\[
\omega=\mbox{const},
\quad
\Lambda(\phi)=-\frac{(\omega+3)(\omega+4)}{2(\omega+1)^2}\,\phi,
\quad
\phi={x^0}^{-1/(\omega+1)} \phi_1(x^1),
\]
\begin{equation}  
	\phi_1{}''=-\frac{\alpha^2(\beta^2-1)}{({x^1}^2-\alpha^2)^2}\,\phi_1-\omega\,\frac{{\phi_1{}'}^2}{\phi_1},
\qquad
\alpha\neq 0,\ \ \ \beta\neq \pm 1.
\label{B11}
\end{equation} 
The solution (\ref{B11}) relates to the Brans-Dicke scalar-tensor theory of gravity.

\subsection{Exact solution \#2 of field equations for model {B1}}

\[
ds^2=\frac{1}{{x^0}^2}\,\left[ 
{dx^0}^2+2\,dx^1dx^3+(x^1-\alpha)^{1-\beta}(x^1+\alpha)^{1+\beta}\,{dx^2}^2
\right],
\quad
\alpha\neq 0,\  
\beta\neq \pm 1,
\]
\begin{equation} 
\omega=-2,
\quad
 \Lambda=-\phi(\phi^2+1),
 \quad
 \phi=\frac{x^0}{x^2}\sqrt{(x^1-\alpha)^{-1+\beta}(x^1+\alpha)^{-1-\beta}}.
\label{B12}
\end{equation} 
In both solutions (\ref{B11}) and (\ref{B12})
for the Riemann curvature tensor $ R_{ijkl} $, the Ricci tensor $ R_{ij} $, the scalar curvature $ R $ and the Weyl tensor $ C_{ijkl} $ we have the following nonzero components:
\begin{equation}  
R_{1313}=-R_{0103}=\frac 1{{x^0}^4},
\quad
R_{1223}=-R_{0202}=\frac{(x^1-\alpha)^{1-\beta}(x^1+\alpha)^{1+\beta}}{{x^0}^4}, 
\end{equation} 
\begin{equation}  
R_{1212}=-\frac{\alpha^2(\beta^2-1)(x^1-\alpha)^{-1-\beta}(x^1+\alpha)^{-1+\beta}}{{x^0}^2},
\end{equation} 
\begin{equation}  
R_{00}=R_{13}=-\frac 3{{x^0}^2},
\qquad
R_{11}=-\frac{\alpha^2(\beta^2-1)}{({x^1}^2-\alpha^2)^2},
\end{equation} 
\begin{equation}  
R_{22}=\frac{3(x^1-\alpha)^{-\beta}(x^1+\alpha)^\beta(\alpha^2-{x^1}^2}{{x^0}^2}, 
\qquad
R=-12, 
\end{equation} 
\begin{equation}  
C_{010}=-\frac{R_{11}}{2{x^0}^2},
\qquad
C_{1212}=\frac 12R_{1212}. 
\end{equation} 
For the solutions (\ref{B11}) and (\ref{B12}) if $ \alpha \ne 0 $ and $ \beta^ 2 \ne 1 $, the Ricci tensor,
 the Riemann curvature tensor, and the Weyl tensor do not vanish.
There are singularities of solutions for $ x^0 = 0 $ and for $ x^1 = \pm \alpha $.

\subsection{Solution of the eikonal equation and the Hamilton-Jacobi equation %of test particle 
of model {B1}}

From the Hamilton-Jacobi equation (\ref{EqHJ}) by the method of complete separation of variables we have
the function $ S $ of the action of the test particle for the model {B1} in the form
($\alpha,\beta,q\ne 0$):
$$
S=m \ln x^0
\pm\left[ \sqrt{m^2+{r}\,{x^0}^2} 
 - m\, \ln \left( m+\sqrt{m^2+{r}\,{x^0}^2} \right)
   \right] - \mbox{}
$$
\begin{equation}
\mbox{}
\frac{2\,\alpha\beta{r}\,x^1+p^2\left[ (x^1-\alpha)/(x^1+\alpha)  \right]^\beta}{4q\alpha\beta}
+px^2+qx^3+ F(p,q,{r}),
%\qquad
%\alpha,\beta,q\ne 0,
\end{equation}
where $ p $, $ q $, $ r $ are the independent constants of the motion of a test particles given by the initial conditions.

The eikonal function from Eq.(\ref{EqEikonal}) in the model {B1} will take the form:
\begin{equation}
\Psi=
r\, x^0
-\frac{2\,\alpha\beta\, {r}^2 x^1+p^2\left[ (x^1-\alpha)/(x^1+\alpha)  \right]^\beta}{4q\alpha\beta}
+px^2+qx^3+ F(p,q,{r}),
\end{equation}
where $ p $, $ q $, $ r $ are independent constants defined by the initial conditions.

\section{Spatially homogeneous wave-like model {B2} type}

The space-time interval for the model {B2} can be written as (see \cite{10}):
\begin{equation}
ds^2=\frac{1}{{x^0}^2}\,\left[ 
{dx^0}^2+2\,dx^1dx^3+{x^1}^{-\alpha}\,{dx^2}^2
\right],
%\qquad
%\alpha \ne 0,-1,
\end{equation}
where $ x^1 $ is an ignored isotropic (wave) variable. %, $ x ^ 3 $ is an ignored isotropic variable.
Independent Killing vector fields in a privileged CS can be selected in the form:
\[
X_0=\partial_3,
\qquad
X_1=\partial_2,
\qquad
X_2=x^0\partial_0+ x^2\partial_2+ 2\,x^3 \partial_3,
\]
\begin{equation}
X_3=\frac{x^0}{2}\,\partial_0+x^1\partial_1+ \frac{\alpha+1}{2}\,x^2\partial_2.
\end{equation}
Killing vectors $ X_1 $, $ X_2 $, $ X_3 $ define a subgroup of spatial homogeneity of the model.
Killing Vectors commutators of  {B2} Model have the form:
\[
[{X_0},{X_1}]=0,
\qquad
[{X_0},{X_2}]=2{X_0},
\qquad
[{X_0},{X_3}]=0,
\]
\begin{equation} 
[{X_1},{X_2}]={X_1},
\qquad
[{X_1},{X_3}]=\frac{\alpha+1}{2}\,{X_1},
\qquad
[{X_2},{X_3}]=0.
\end{equation}

For $ {\alpha} = - 1 $, this space admits an additional Killing commuting vector and degenerates into a space with one non-ignored variable.

The condition of spatial homogeneity of the model imposes restrictions on the range of allowed values ​​of variables.
It follows from the form of the metric determinant that $ x^1> 0 $, and the condition of the spatial homogeneity of the model leads to restrictions on the allowed range of variables.
The homogeneous space in the used coordinate system belongs to the region $ x^3 <0 $ and $ {x^0}^2 > 2 \,
 | x^1x^3 | $.
 
The scalar curvature of the model is constant and negative,
the components of the Weil tensor are proportional to the expression $ C_{ijkl} \sim \alpha \, (\alpha + 2) $.

This spatially homogeneous space-time model is of type {III} according to the Bianchi classification and has type N according to the Petrov classification.

If the value of the parameter $ {\alpha} = 0 $, the metric of the model degenerates -
in a privileged CS, it depends on only one variable $ x^0 $, and the model becomes conformally flat.

The field equations (\ref{EqField}) for the model {B2} take the form:
\[
\Lambda  =  \Bigl[\phi_0^2 \phi_1 \phi_3 \omega {x^0}^2     {x^1}^\alpha \phi_2{}'^2 -    
\phi_2^2\,
\Bigl(
\phi_1 \phi_3 (6 \phi_0^2 - 6 \phi_0 {x^0}    
      \phi_0{}' - \omega          {x^0}^2 \phi_0{}'^2)+      
\mbox{}
\]
\begin{equation}  
2 \phi_0^2 (3 + 2 \omega) {x^0}^2       \phi_1{}' \phi_3{}'
\Bigr)
\Bigr]
/(2 \phi_0 \phi_2) 
\label{EqFieldB21}
\end{equation} 
\begin{equation}  
\phi_2{}''  =   -\frac{\omega \phi_2{}'^2}{\phi_2}     + \phi_2 {x^1}^{(-1 - \alpha)}
\Bigl(
\alpha \phi_1 +      2 (1 + \omega) {x^1}       \phi_1{}'
\Bigr) 
\phi_3{}'/(2 \phi_1 \phi_3)
 \end{equation} 
\begin{equation}  
\phi_0{}''  = - \frac{2 \phi_0{}'}{{x^0}} -   \frac{\omega \phi_0{}'^2}{\phi_0}    + \frac{\phi_0 \phi_1{}' \phi_3{}'}{\phi_1 \phi_3} +
 \frac{\phi_0 \omega     \phi_1{}' \phi_3{}'}{\phi_1 \phi_3} 
 \end{equation} 
\begin{equation}  \phi_1{}''  =  -(\alpha (2 + \alpha) \phi_1)/(4 {x^1}^2) -   \frac{\omega \phi_1{}'^2}{\phi_1} \end{equation} 
\begin{equation}  (\alpha \phi_1 +    2 (1 + \omega) {x^1}     \phi_1{}') \phi_2{}'=0 \end{equation} 
\begin{equation}  (\phi_0 + (1 + \omega)     {x^0} \phi_0{}') \phi_1{}'=0 \end{equation} 
\begin{equation}  (\phi_0 + (1 + \omega)     {x^0} \phi_0{}') \phi_2{}'=0 \end{equation} 
\begin{equation}  (\phi_0 + (1 + \omega)     {x^0} \phi_0{}') \phi_3{}'=0 \end{equation} 
\begin{equation}  \omega     \phi_3{}'^2 + \phi_3 \phi_3{}''=0 \end{equation} 
\begin{equation}  (\omega+1)\phi_2{}'\phi_3{}'=0 \end{equation} 
The scalar equation (\ref{EqScalar}) for the model {B2} has the form:
\[
0=
 -2\Lambda' \phi_0 \phi_1 \phi_2 \phi_3 +  4 \Lambda + \omega' \phi_1 \phi_3 {x^0}^2   (\phi_1 \phi_3 (\phi_2^2 \phi_0{}'^2 +      \phi_0^2 {x^1}^\alpha \phi_2{}'^2) +    2 \phi_0^2 \phi_2^2 \phi_1{}' \phi_3{}') +  
 \mbox{}
\]
\[
(3 + 2 \omega) {x^0}   
\Bigl[
  4 \phi_0 \phi_2 {x^0} {x^1} \phi_1{}'  \phi_3{}' + 
 \phi_1      
 \Bigl(
  - \phi_2
\bigl(
\alpha \phi_0 {x^0} \phi_3{}' +  2 \phi_3 {x^1} (2 \phi_0{}' - {x^0} \phi_0{}'')
\bigl) +
  \mbox{}
 \]
\begin{equation}  
  2 \phi_0 \phi_3 {x^0} {x^1}^{(1 + \alpha)}\phi_2{}''
\Bigl)
\Bigr]
  /(2 {x^1}) .
\label{EqScalarB2}
\end{equation} 
Integrating the equations (\ref{EqFieldB21})-(\ref{EqScalarB2}) gives four exact solutions for the {B2} model.
We list them below.

\subsection{Exact solution \#1 of field equations for model {B2}}

\[
ds^2=\frac{1}{{x^0}^2}\,\left[ 
{dx^0}^2+2\,dx^1dx^3+{x^1}^{2}\,{dx^2}^2
\right],
\]
\begin{equation} 
\omega=-\frac{\phi_0\phi_0{}''}{{\phi_0{}'}^2}-2\frac{\phi_0}{\phi_0{}'x^0},
\qquad
\Lambda=-3\phi_0+2\phi_0{}'x^0-\frac 12\phi_0{}''{x^0}^2,
\label{SolB21}
\end{equation} 
where the scalar field $ \phi = \phi_0 (x^0) $ is an arbitrary function of $ x^0 $.

Note that with a power-law dependence of the scalar field on $ x^0 $ of the form
$ \phi = \phi_0 = \alpha {x^0}^\beta $ ($ \alpha, \beta $ - const), this solution leads to the Brans-Dicke scalar-tensor theory of gravity.

The solution (\ref{SolB21}) is conformally flat (the Weyl tensor is zero), but the Ricci tensor, scalar curvature, and the Riemann curvature tensor do not vanish.

\subsection{Exact solution \#2 of field equations for model {B2}}

\[
ds^2=\frac{1}{{x^0}^2}\,\left[ 
{dx^0}^2+2\,dx^1dx^3+{x^1}^{2}\,{dx^2}^2
\right],
\]
\[
\omega=\mbox{const},
\qquad
 \Lambda=\frac 1{2(\omega+1)^2}\,\phi \left(
 \omega\,\phi^{- 2/(\omega+1)}+(\omega+3)(\omega+4)
 \right),
\]
\begin{equation}  
\phi=\left(
{x^1}{x^2}/{x^0}
\right)^{1/(\omega+1)}.
 \label{SolB22}
\end{equation} 
In both solutions (\ref{SolB21}) and (\ref{SolB22}) the Weyl tensor $ C_{ijkl}=0 $,
for the Riemann curvature tensor $ R_{ijkl} $, the Ricci tensor $ R_{ij} $, the scalar curvature $ R $  we have the following nonzero components:
\begin{equation}  
R_{1313}=-R_{0103}=\frac 1{{x^0}^4},\ \ \ R_{1223}=-R_{0202}=\frac{{x^1}^2}{{x^0}^4}, 
\end{equation} 
\begin{equation} 
 R_{00}=R_{13}=-\frac 3{{x^0}^2},\ \ \ R_{22}=-\frac{3{x^1}^2}{{x^0}^2},\ \ \ R=-12. 
\end{equation} 
The models (\ref{SolB21}) and (\ref{SolB22}) are conformally flat (the Weyl tensor is zero), but the Ricci tensor, scalar curvature, and the Riemann curvature tensor do not vanish. 

\subsection{Exact solution \#3 of field equations for model {B2}}

\[
ds^2=\frac{1}{{x^0}^2}\,\left[ 
{dx^0}^2+2\,dx^1dx^3+{x^1}^{-\alpha}\,{dx^2}^2
\right],
%\qquad
%\alpha \ne 0,-1,
\]
\begin{equation}  
\omega=-2,
\qquad
\Lambda=-\phi(\phi^2+1),
\qquad
\phi=x^0{x^1}^{\alpha/2}/x^2.
\label{SolB23}
\end{equation} 

\subsection{Exact solution \#4 of field equations for model {B2}}

\[
ds^2=\frac{1}{{x^0}^2}\,\left[ 
{dx^0}^2+2\,dx^1dx^3+{x^1}^{-\alpha}\,{dx^2}^2
\right],
%\qquad
%\alpha \ne 0,-1,
\]
\[
\omega=\mbox{const},
\qquad
\Lambda=-\frac{(\omega+3)(\omega+4)}{2\,(\omega+1)^2}\,\phi,
\]
\[
{\phi}=p\left[
\left({{x^1}^{(1+\sigma)/2}+q\, {x^1}^{(1-\sigma)/2}}\right)/{x^0}
\right]^{1/(1+\omega)},
\qquad
p, q - \mbox{const},
\]
\begin{equation}
\sigma=\pm \sqrt{1-\alpha^2(2+\alpha)(1+\omega)}.
\label{SolB24}
\end{equation}
The solution (\ref{SolB24}) relates to the Brans-Dicke scalar-tensor theory of gravity.

In both solutions (\ref{SolB23}) and (\ref{SolB24})
for the Riemann curvature tensor $ R_{ijkl} $, the Ricci tensor $ R_{ij} $, the scalar curvature $ R $ and the Weyl tensor $ C_{ijkl} $ we have the following nonzero components:
\[
R_{1313}=-R_{0103}=\frac 1{{x^0}^4},\ \ \ R_{1223}=-R_{0202}=\frac{{x^1}^{-\alpha}}{{x^0}^4},\ \ \ R_{1212}=-\frac{\alpha(\alpha+2)}{4{x^0}^2{x^1}^{\alpha+2}}, 
\]
\begin{equation} 
 R_{00}=R_{13}=-\frac 3{{x^0}^2},\ \ \ R_{22}=-\frac{3{x^1}^{-\alpha}}{{x^0}^2},\ \ \ R_{11}=-\frac{\alpha(\alpha+2)}{4{x^1}^2},\ \ \ R=-12, \end{equation} 
\begin{equation}  C_{010}=-\frac{R_{11}}{2{x^0}^2},\ \ \ C_{1212}=\frac 12R_{1212}.
 \end{equation} 
In the solutions (\ref{SolB23}) and (\ref{SolB24}) for $ \alpha = -2 $ or $ \alpha = 0 $, the Weyl tensor vanishes (conformally flat spaces), but the Ricci tensor, scalar curvature and the Riemann curvature tensor does not vanish. %Models have a singularity at the origin \ mbox {for $ x ^ 0 x ^ 1 = 0 $}.

\subsection{Solution of the eikonal equation and the Hamilton-Jacobi equation of the model {B2}}

Let us present for the model {B2} the result of integrating the Hamilton-Jacobi equation (\ref{EqHJ}) by the method of complete separation of variables. The function $ S $ of the action of the test particle for the model {B2} can be written in the following form ($ \alpha q \ne 0 $):
\[
S=m \ln {x^0}+\sqrt{m^2-{r} {x^0}^2}-m \ln \left(m+\sqrt{m^2-{r} {x^0}^2}\right)
+\mbox{}
\]
\begin{equation}
\frac{2 \alpha  {r} {x^1}-p^2 \left({x^1}\right)^{2 \alpha } }{4 \alpha  q}
+ p   {x^2}+q {x^3}+ F(p,q,{r}),
\end{equation}
where $ p $, $ q $, $ r $ are the independent constant motions of the test particles given by the initial conditions.

Solution of the eikonal equation (\ref{EqEikonal}) for the {B2} model %  for $ q \ ne 0 $
has the form:
% ($q\ne 0$):
\begin{equation}
\Psi={r}x^0-\frac{1}{2q}\left( {r}^2x^1+\frac{p^2}{2\alpha}\,{x^1}^{2\alpha}\right)+ p x^2+q x^3 +F(p,q,{r}),
\quad
\alpha\,q \ne 0,
\end{equation}
where $ p $, $ q $, $ r $ are independent constants defined by the initial conditions.

\section{Conclusion}

Six exact solutions are obtained in the general scalar-tensor theory of gravity related to spatially homogeneous wave-like models of the Universe. Wave-like space-time models allow the existence of privileged coordinate systems, where the eikonal equation and the Hamilton-Jacobi equation of test particles allow complete separation of variables with separation of isotropic (wave) variables, on which the space-time metric  depends (non-ignored variables).
For all the solutions found, an explicit form of the scalar field and two functions of the scalar field that are part of the general scalar-tensor theory of gravity is obtained.
For the models under consideration, an explicit form of the eikonal function for radiation and the action function for test particles is obtained, which allows us to completely describe the motion of test particles and light rays.
The obtained solutions are of type~III according to the classification \mbox{Bianchi} and type~N according to the classification of \mbox{Petrov}.

Wave-like spatially homogeneous space-time models can describe aperiodic primordial gravitational waves of the Universe.

\section*{Acknowledgments}

The reported study was funded by RFBR, project number N~20-01-00389~A.

\end{document}